**Angelo Raffaele Meo**
**Accademia delle Scienze di Torino**

# ON THE P VS NP QUESTION: A PROOF OF INEQUALITY

**Summary**

The analysis discussed in this paper is based on a well-known NP-complete problem which is called "satisfiability problem or SAT". From SAT a new NP-complete problem is derived, which is described by a Boolean function called "core function". In this paper it is proved that the cost of the minimal implementation of core function increases with **n** exponentially. Since the synthesis of core function is an NP-complete problem, this result is equivalent to proving that **P** and **NP** do not coincide.

## 1. INTRODUCTION

A brief description of the definitions and properties well known among the scientists of modern computational complexity theory which will be made reference to, is presented in this section.

**P** denotes the class of all the decision problems which can be solved in polynomial time.

**NP** denotes the class of all the decision problems **f** satisfying the property that the function **check(f)** analyzing a witness of the decision problem is polynomial time decidable.

"**P=NP?**", or, in other terms, "Is **P** a proper subset of **NP**?", is one of the most important open questions in modern computational complexity theory.

A decision problem **C** in **NP** is **NP-complete** if it is in **NP** and if every other problem **L** in **NP** is reducible to it, in the sense that there is a polynomial time algorithm which transforms instances of **L** into instances of **C** producing the same values.

The importance of NP-completeness derives from the fact that, if we find a polynomial time algorithm for just one **NP-complete** problem, then we can construct polynomial time algorithms for all the problems in **NP** and, conversely, if any single **NP-complete** problem does not have a polynomial time algorithm, than no **NP-complete** problem has a polynomial time solution.

The analysis discussed in this paper will be based on a well-known **NP-complete** problem which is called "satisfiability problem or **SAT**".

Given a Boolean expression containing only the names of a set of variables (some of which may be complemented), the operators **AND**, **OR** and **NOT**, and parentheses, is there an assignment of **TRUE** and **FALSE** values to the variables which makes the entire expression **TRUE**?

It is well known that the problem remains **NP-complete** also when all the expressions are written in "conjunctive normal form" with **3** variables per clause (problem **3SAT)**. In this case, the analyzed expressions will be of the type:

$$
\begin{aligned}
&\mathbf{F=(}\ x_{11}\ \mathbf{OR}\ x_{12}\ \mathbf{OR}\ x_{13}\ \mathbf{)\ AND}\\
&\mathbf{(}\ x_{21}\ \mathbf{OR}\ x_{22}\ \mathbf{OR}\ x_{23}\ \mathbf{)\ AND}\\
&\ldots\ldots\ldots\ \mathbf{AND}\ \ldots\ldots\ldots\\
&\mathbf{(}\ x_{t1}\ \mathbf{OR}\ x_{t2}\ \mathbf{OR}\ x_{t3}\ \mathbf{)}
\end{aligned}
\qquad (1)
$$

where:

**t** is the number of clauses or triplets;

each $\mathbf{x_{ij}}$ is a variable in complemented or uncomplemented form;

each variable can appear multiple times in the expression.

If the deterministic Turing machine is assumed as the computational model, with **{0,1,b}** as its set of input symbols, the input data appearing on the tape at the beginning of computation can represent the data of expression **(1)** in the following way:

**b b <binary code of number of variables> <separator>b**

or

$$
\begin{aligned}
&s_{11}\, n_{111}\, n_{112}\, n_{113} \ldots\ldots n_{11m}\mathbf{b}\\
&s_{12}\, n_{121}\, n_{122}\, n_{123} \ldots\ldots n_{12m}\mathbf{b}\\
&s_{13}\, n_{131}\, n_{132}\, n_{133} \ldots\ldots n_{13m}\mathbf{b}\\
&s_{21}\, n_{211}\, n_{212}\, n_{113} \ldots\ldots n_{21m}\mathbf{b}\\
&\ldots\ldots\ldots\\
&s_{t3}\, n_{t31}\, n_{t32}\, n_{t33} \ldots\ldots n_{t3m}\mathbf{b}
\end{aligned}
\qquad (2)
$$

where:

**b** is the blank symbol;

**t** is the number of triplets;

$\mathbf{s_{ij}}$ denotes the sign of variable $\mathbf{x_{ij}}$

(with $\mathbf{s_{ij} = 1}$ denoting that $\mathbf{x_{ij}}$ is preceded by operator **NOT**);

$\mathbf{n_{ijk}}$ denotes the **k-th** component of the binary code $\mathbf{<n_{ij1}\ n_{ij2}\ \ldots\ n_{ijm}>}$ representing the name of variable $\mathbf{x_{ij}}$ ;

the binary code of the number $\mathbf{n_v}$ of variables is needed in order to determine the number **m** of binary digits necessary to represent the names of variables according the rule

**m** = *minimum integer not less than* $\mathbf{\log_2 n_v}$

Notice that, by neglecting the bits of the binary code of the number of variables and the bits of the separator, the number of input bits on the tape will be

$$t \cdot 3 \cdot (1 + \textit{minimum integer not smaller than } \log_2 (3 \cdot t)) \quad (3)$$

since the maximum value of the number of variables is **3·t**, where **t** plays the role which is usually called "**n**".

The properties of Turing machines processing the bit string described by **(2)** will be analyzed in this paper with reference to a family **{$C_n$}** of Boolean circuits, where **$C_n$** has **n** binary inputs and produces the same binary output as the corresponding Turing machine.

The equivalence between a deterministic Turing machine **M** processing some input **x** belonging to **$\{0,1\}^n$** and an **n**-input Boolean circuit **$C_n$** is well known. It is also known that the number of gates, or **AND, OR, NOT** operators,i appearing in circuit **$C_n$**, is polynomial in the running time of the corresponding Turing machine.

## 2. THE CORE FUNCTION

In the case of satisfiability problem with **3** variables for clause, Boolean circuit **$C_n$** has **n** **(=t)** sets of inputs which the binary data described in **(2)** are applied to. (Of course, the binary code of the number of variables and the separator are not needed). The output of **$C_n$** (with **n=t**) will take the value **TRUE** when, and only when, there is an assignment of values **TRUE** and **FALSE** to variables making expression **(1) TRUE**.

In order to simplify analysis, circuit **$C_n$** will be decomposed into two processing layers as shown in **Fig. 1**, where , as usual, the number **t** of triplets plays the role of symbol **n** in the standard analysis of complexity theory.

In the following analysis, we shall use the symbol **t** when it's necessary to remember the number of triplets and **n** in the other cases.

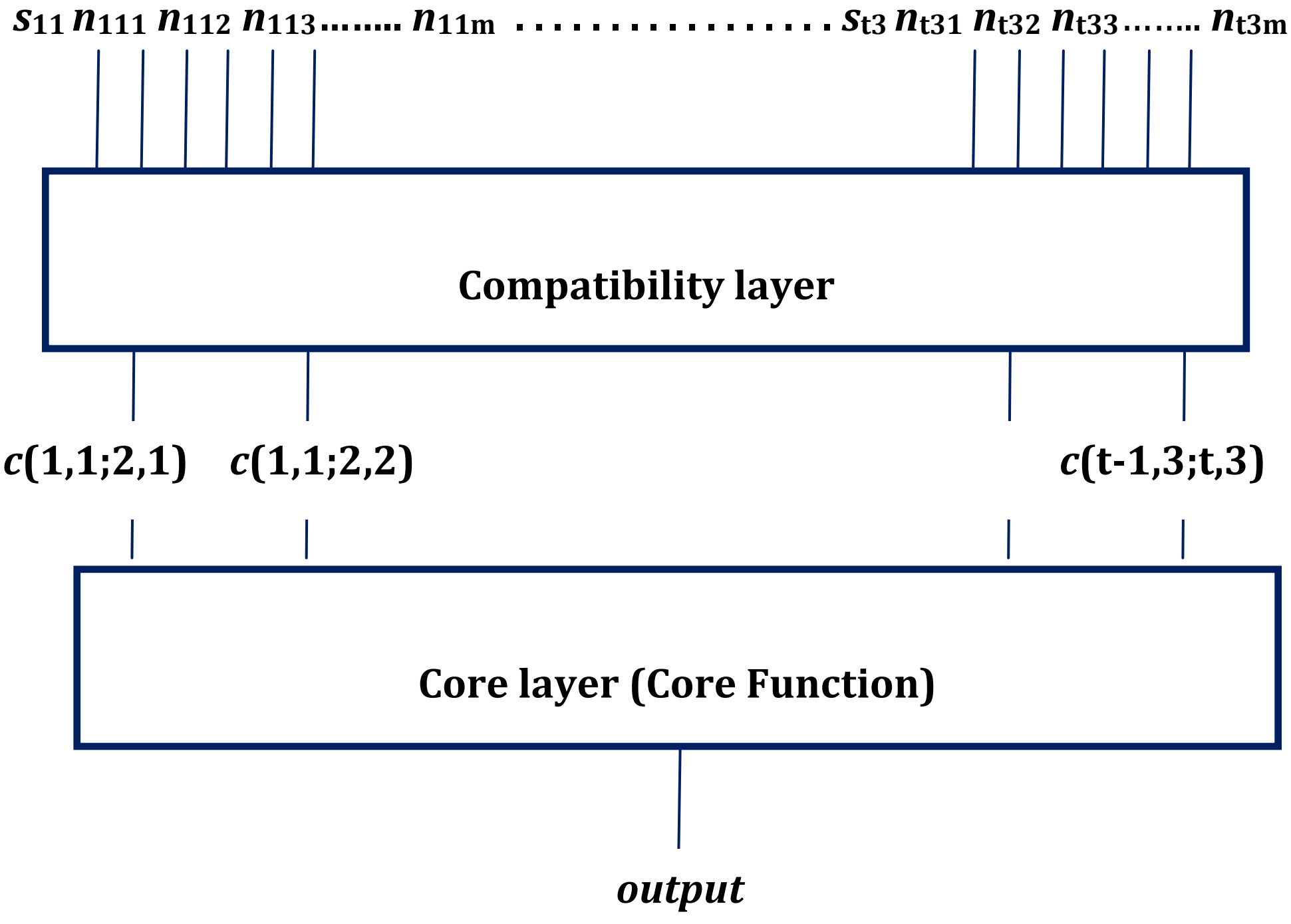

**Fig. 1**

**Decomposition of Boolean circuit $C_n$ into compatibility layer and core layer**

A variable **j** of triplet **i** will be defined as "**compatible**" with variable **k** of triplet **h** when, and only when either

- the sign $\mathbf{s_{ij}}$ of the former variable is equal to the sign $\mathbf{s_{hk}}$ of the latter,

or

- the name **<**$\mathbf{n_{ij1}\ n_{ij2}\ ...n_{ijm}}$**>** of the former is different from the name **<**$\mathbf{n_{hk1}\ n_{hk2}\ ...n_{hkm}}$**>** of the latter.

From that definition it follows that two "**not compatible**" variables have different signs and the same name; therefore, their **AND** are identically **FALSE**.

The compatibility layer is composed of **3·t·(3·t-3)/2** identical cells, one for each pair of variables belonging to different triplets.

As shown in **Fig. 2**, the inputs of a cell will be the sign $\mathbf{s_{ij}}$ and the name **<**$\mathbf{n_{ij1}\ n_{ij2}\ ...n_{ijm}}$**>** of variable **j** of triplet **i**, and the sign $\mathbf{s_{hk}}$ and the name **<**$\mathbf{n_{hk1}\ n_{hk2}\ ...n_{hkm}}$**>** of variable **k** of triplet **h**. The output of the same cell **c(i,j;h,k)** will be **TRUE** when, and only when, the two variables are compatible between themselves.

**c(i,j;h,k)=** → **TRUE** $\Leftrightarrow$ $x_{ij}$ **is compatible with** $x_{hk}$

→ **FALSE** $\Leftrightarrow$ $x_{ij}$ **is not compatible with** $x_{hk}$

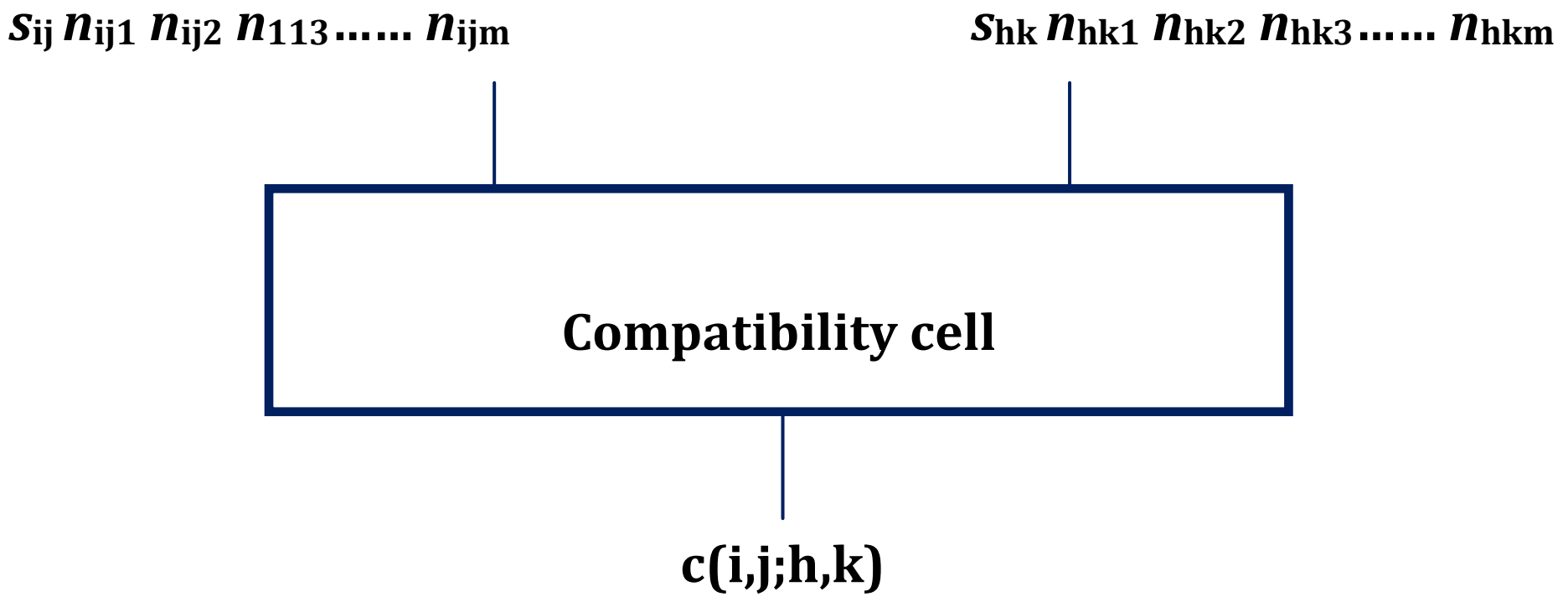

**Fig. 2**

**Compatibility Cell**

Variable **c(i,j;h,k)** will be called a compatibility variable or simply a compatibility.

The core layer processes only the **9·t·(t-1)/2** compatibility variables **c(i,j;h,k)** and produces the global result of computation.

As the circuit $\mathbf{C_n}$, also the global Boolean function implemented by $\mathbf{C_n}$ may be decomposed into two layers of functions. At the compatibility layer, the function implemented by a cell may be written as follows (by using the symbols ∗, **+**, and **!** for representing **AND, OR** and **NOT** operators, respectively):

$$c(i,j;h,k) = s_{ij}*s_{hk} + !s_{ij}*!s_{hk} + \quad \textit{(equal sign)}$$

$$\left.\begin{aligned} &+ n_{ij1}*!n_{hk1} + !n_{ij1}*n_{hk1} + \\ &+ n_{ij2}*!n_{hk2} + !n_{ij2}*n_{hk2} + \\ &\ldots\ldots\ldots \\ &+ n_{ijm}*!n_{hkm} + !n_{ijm}*n_{hkm} \end{aligned}\right\} \textit{(at least one bit in the variables names different)} \qquad \textbf{(4)}$$

The Boolean function implemented by the core layer will be called the "**Core Function**" of order **t**, where **t** is the number of triplets. It will be denoted with the symbol **CF**(**t**) (or **CF(n)**). The core function can be determined by proceeding as follows.

Consider one selection of variables appearing in **(1)**, one and only one for each triplet, for all the triplets. Let

$$<1i_1>, <2i_2>, \ldots, <ti_t> \qquad \textbf{(5)}$$

with $i_1, i_2, \ldots, i_t \in \{1, 2, 3\}$

be the indexes **<number of triplet, number of variable in the triplet>** of the selected variables. They will be called "characteristic indexes". Let $\Pi^k$ be the product of all the compatibility variables relative to the **k-th** of selections **(5)**:

$$\Pi^k = c(1,i_1; 2,i_2)*c(1,i_1; 3,i_3)*\ldots \ldots*c(t-1,i_{t-1}; t, i_t) \qquad \textbf{(6)}$$

The core function can be defined as the sum

$$\Sigma_k\Pi^k \qquad \textbf{(7)}$$

of the products **(6)** relative to all the selections **(5)**.

For example, in the case of **CF(3)**, the core function can be defined as follows:

$$\begin{aligned} CF(3) = &c(1,1;2,1)*c(1,1;3,1)*c(2,1;3,1) + \\ &c(1,1;2,1)*c(1,1;3,2)*c(2,1;3,2) + \\ &c(1,1;2,1)*c(1,1;3,3)*c(2,1;3,3) + \\ &c(1,1;2,2)*c(1,1;3,1)*c(2,2;3,1) + \\ &\ldots\text{(other 22 products)}\ldots + \\ &c(1,3;2,3)*c(1,3;3,3)*c(2,3;3,3) \end{aligned} \qquad \textbf{(8)}$$

It is easy to prove that there is an assignment of value **TRUE** or **FALSE** to variables appearing in Eq. **(1)** which make the value of **(1)** equal to **TRUE** when, and only when, the core function takes the value **TRUE.**

Notice that the processing work of the cell of **Fig. 2** increases as a polynomial function **P(t)** of the number of the variables since the increment of the length of the code of the name is logarithmic. Therefore, the total processing work of the compatibility layer increases as:

$$9 \cdot t \cdot (t-1) \cdot P(t)$$

where $9 \cdot t \cdot (t-1)/2$ is the total number of the compatibility cells.

Besides, the problem solved by the core layer is clearly in **NP**, because it is easy to verify a witness solution. It follows that, since the compatibility layer polynomially reduces an NP-complete problem **(3SAT)** to the problem solved by the core layer, the core function describes a new NP-complete problem.

Some interesting properties of core function have been discussed in ref. (23).

## 3. A THEOREM OF BOOLEAN MONOTONIC FUNCTIONS

Let $\mathbf{f(x_1,x_2, ..., x_t)}$ be an isotonic Boolean function, that is a Boolean function which can be implemented with only **AND** and **OR** gates, applied to uncomplemented literals $\mathbf{x_1, x_2, ..., x_t}$. It was believed that the minimum cost implementation of $\mathbf{f(x_1,x_2,...,x_t)}$ always contains only **OR** and **AND** gates, but A.Razborov proved that there are isotonic functions whose minimum cost implementation contains also **NOT** gates (see ref. (**8**) ).

However, there is on upper bound on the comparison of the costs of the minimum cost implementations with and without **NOT** gates. It is specified by the following theorem.

### 3.1. <u>Theorem</u>

Let $\mathbf{I_{min}}$ be one of the minimum cost implementations of the isotonic Boolean function $\mathbf{f(x_1, x_2,...,x_t)}$, the cost being defined as the total number of **AND**, **OR** or **NOT** gates. Let $\mathbf{C_{min}}$ be the cost of $\mathbf{I_{min}}$.

There exists always an implementation **J** of **f** containing only **AND** and **OR** gates such that

$$\mathbf{cost\ (J) <= 2 \cdot C_{min} + t}$$

In order to prove this theorem, let us divide the gates of implementation $\mathbf{I_{min}}$ of **f** into different levels.

At level **1** we place the gates all inputs of which coincide with the complemented or uncomplemented input variables $\mathbf{x_i}$ or $\mathbf{!x_i}$ (where $\mathbf{!x_i}$ denotes the complement of variable $\mathbf{x_i}$) .

Level **2** contains the gates whose inputs coincide with input variables or outputs of level **1** gates.

In general terms, level **q** contains the gates whose inputs coincide with input variables or outputs of levels less than **q**.

We can transform $\mathbf{I_{min}}$ into **J** by deleting **NOT** gates and adding new **AND** or **OR** gates as follows.

We start from level **1**.

For any level **1 AND** gate we add an **OR** gate whose inputs are the complements of the inputs of the considered **AND** gate **(Fig. 3)**. Similarly, for any level **1 OR** gate we add an **AND** gate whose inputs are the complements of the corresponding **OR** gate.

By virtue of such operations, for any output **u** of the level **1** gates a new node will be available in the new circuit we are generating whose value will be **!u.**

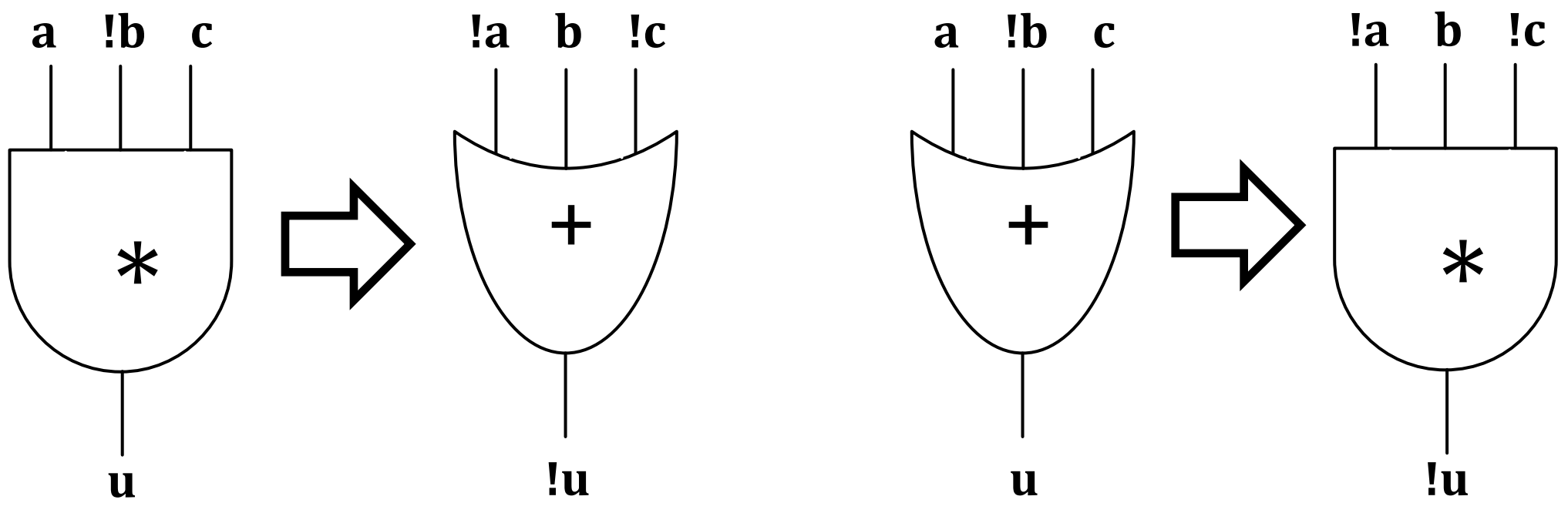

**Fig. 3**

**The transformation of gates of level 1**

As a second step of processing, for any level **2 AND** gate of implementation $\mathbf{I_{min}}$ we shall add an **OR** gate whose inputs are the complements of the inputs of the corresponding **AND** gate, in both the cases in which these inputs coincide with input variables of **f** or with outputs of level **1** gates (**Fig. 4**).

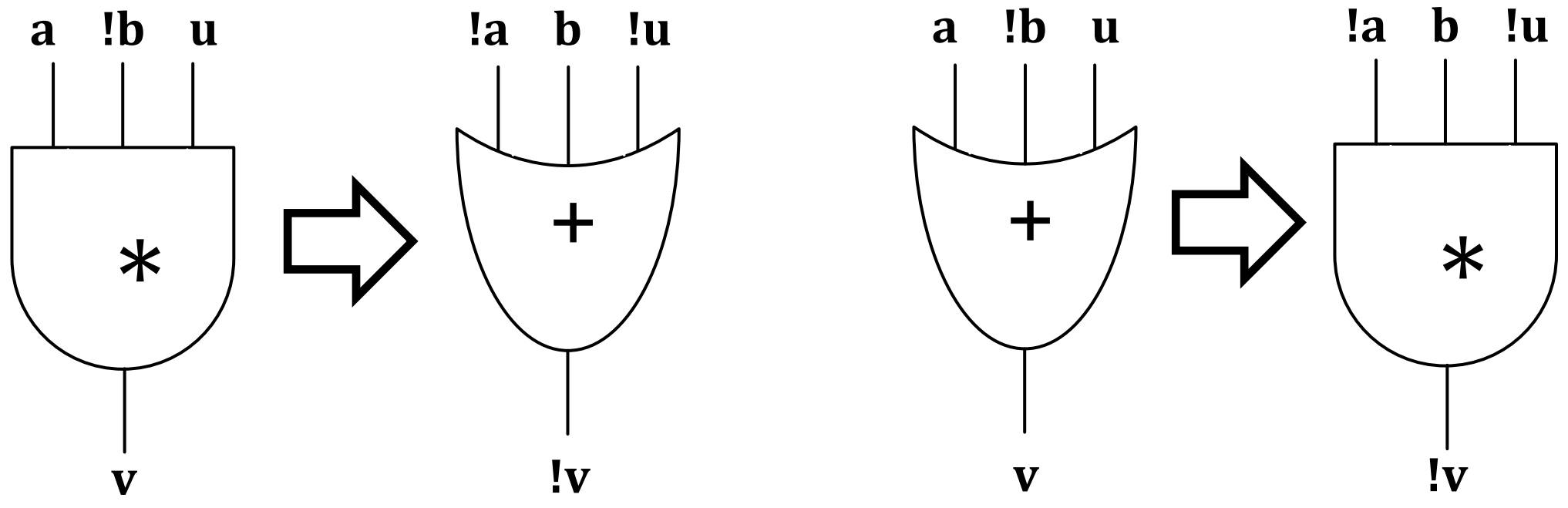

**Fig. 4**

**The transformation of gates of level 2**

A similar transformation will be applied to all level **2 OR** gates.

As an example, the two level subnetwork of **Fig. 5** will be transformed into the subnetwork of Fig. **6**. Notice that at the outputs of **J** not only the outputs **v** and **w** of $\mathbf{I_{min}}$ will be available, but also their complements !**v** and !**w**.

The preceding operations will be applied to all the levels of implementation $\mathbf{I_{min}}$, in the order of increasing levels. It is apparent that, if for any input variable $\mathbf{x_i}$ also $\mathbf{!x_i}$ is available, the number of gates of **J** is less than twice the number of gates of $\mathbf{I_{min}}$.

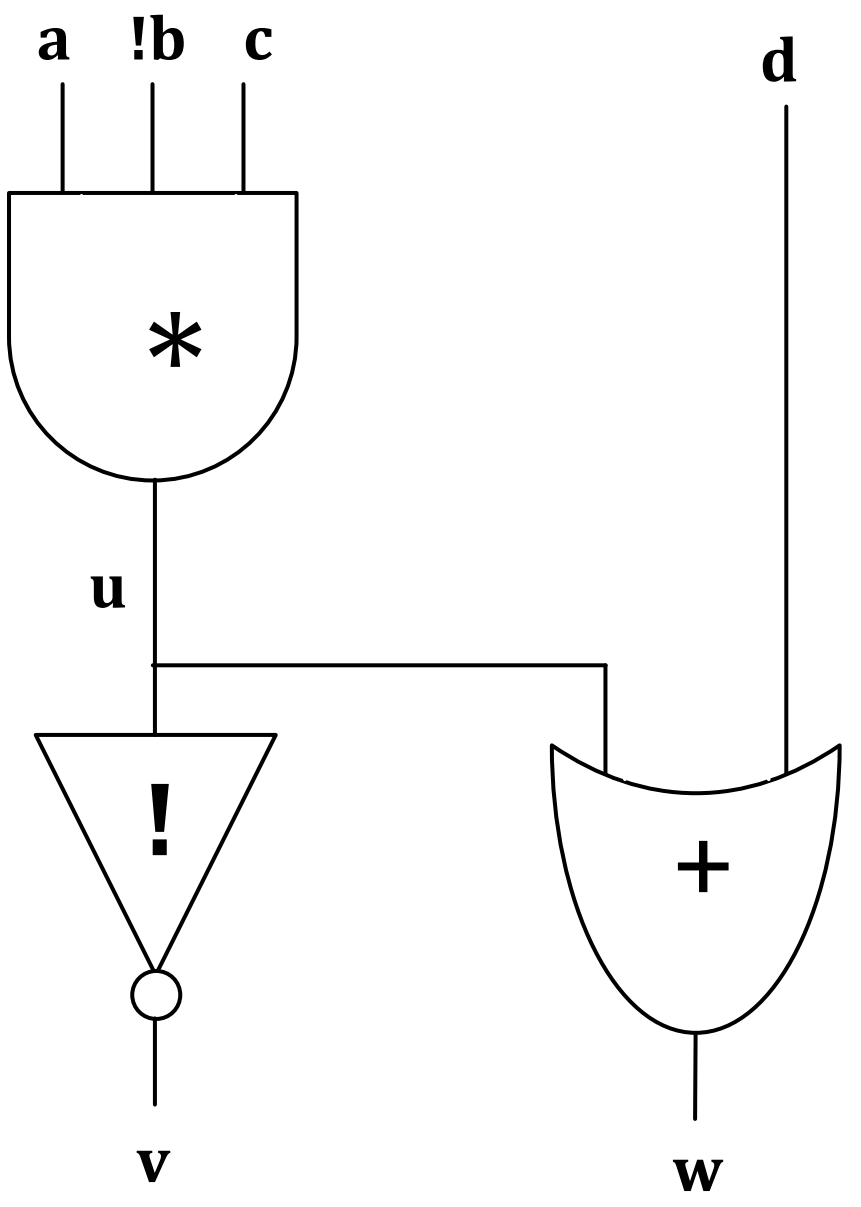

**Fig. 5**

**A two level subnetwork**

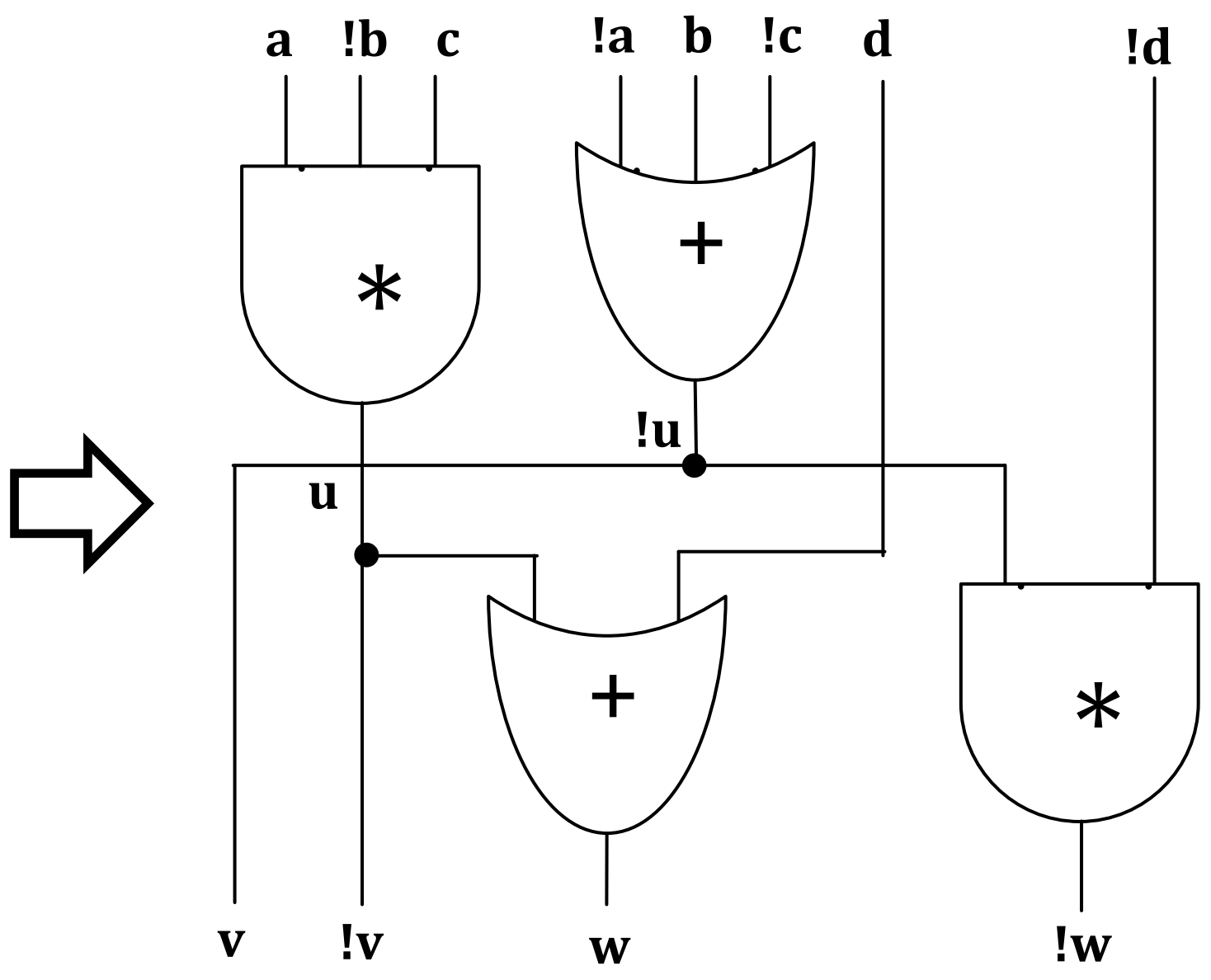

**Fig. 6**

**The transformation of the subnetwork**

At level **0**, before the gates of **Fig. 6**, **t NOT** gates might be necessary to generate the complemented input variables **!$x_i$**. Therefore, **t** has been added in the statement of the theorem.

This theorem will be very important in order to simplify the analysis of core function circuits.

## 4. PROPERTIES OF CORE FUNCTION

It is easy to prove the following properties of core function.

### 4.1. <u>Property 1</u>

Core function is totally isotone.

### 4.2. <u>Property 2</u>

Any product **(6)** is a prime implicant of core function (that is, a product of compatibilities ("**PoC**") which implies core function and no other term of it).

### 4.3. <u>Property 3</u>

Since the different selections of each of variables **(5)** are **3**, the number of prime implicants of the core function is equal to **$3^t$**. Each of these prime implicants is essential (that is, it does not imply a sum of other prime implicants) and it is the product of **$t \cdot (t-1)/2$** compatibilities.

## 5. PRODUCTS OF COMPATIBILITIES

In the next section, reference will be made to the following definitions.

### 5.1. Definition of spurious compatibilities pair

A pair of compatibility variables {**c(h,k;l,m), c(p,q;r,s)**} is defined as a spurious pair **if**

**( h = p and k ≠ q )**

**or** **( h = r and k ≠ s )**

**or** **( l = p and m ≠ q )**

**or** **( l = r and m ≠ s )**

In a graphic scheme:

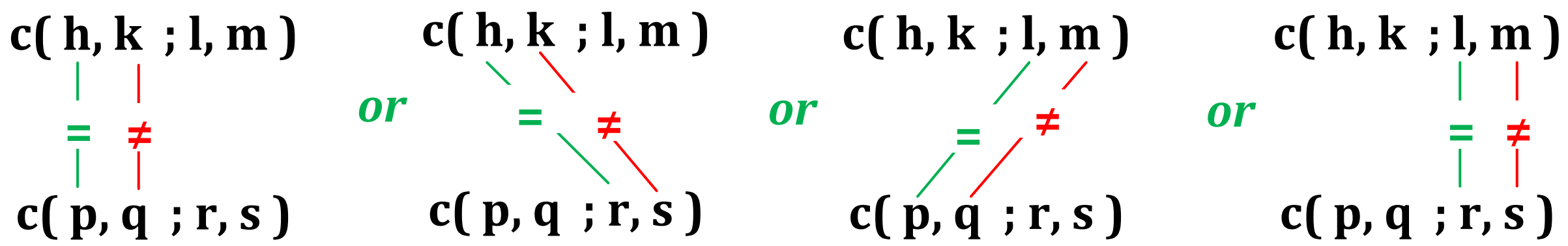

For example, the pair {**c(1,1;2,1), c(1,2;3,1)**} is a spurious pair since the triplet **1** is associated to two different indexes of variables (**1** and **2**).

### 5.2. Definition of spurious products of compatibilities

A spurious product of compatibilities (spurious **PoC**) is a product of compatibility variables containing the elements of one or more than one spurious pair.

For example, the **PoC**

**c(1,1;2,1)∗c(1,2;3,1)∗c(2,1;3,1)**

is a spurious **PoC** since it contains the elements of the spurious pair

**{c(1,1;2,1), c(1,2;3,1)}**

### 5.3. Definition of impure products of compatibilities

A **PoC** containing one or more complemented variables will be defined as an impure **PoC**. In particular a term **T** of **CF** (that is, a **PoC** implying **CF**) that contains one or more complemented variables, will be defined as an impure term.

### 5.4. Definition of core of a PoC

The product of all the uncomplemented variables of **T** will be defined as the core of **T.**

### 5.5. Definition of mark

Consider a not spurious subset of compatibilities satisfying the property that each of the indexes of triplet appears at least once in some variable. The product of the variables of such a subset will defined as a "mark" of the prime implicant of which it contains a subset of compatibilities.

For example, in the case of **CF** (**4**), the **PoC**

**M = c(1,a;2,b)∗c(1,a;3,c)∗c(1,a;4,d)** **(9)**

(where **a, b, c, d** are elements of **{1,2,3}**)

is a mark of the prime implicant

**P= c(1,a;2,b)∗c(1,a;3,c)∗c(1,a;4,d)∗c(2,b;3,c)∗c(2,b;4,d)∗c(3,c;4,d)** **(10)**

since all the indexes of triplet appear at least once in **(9)**.

## 5.6. DEFINITION OF SPURIUS MARK

A spurious **PoC** in which all the indexes of triplet appear at least once will be called a "spurious mark". Notice that a spurious mark may be the mark of more than one prime implicant. For the example, in the case of **CF(3),**

**c(1,1;2,1)∗c(1,1;3,1)∗c(1,1;2,2)**

is a spurious mark of both the prime implicants

**c(1,1;2,1)∗c(1,1;3,1)∗c(2,1;3,1)**

and

**c(1,1;2,2)∗c(1,1;3,1)∗c(2,2;3,1)**

An impure **PoC** whose core is a (possibly spurious) mark will be a defined as a (possibly spurious) impure mark.

## 5.7. DEFINITION OF EXTENDED PRIME IMPLICANTS

A term **T** of core function, that is, an implicant of core function (a product of literals implying core function), contains all the uncomplemented literals of a prime implicant. Therefore, it may be defined as an "extended prime implicant" (only) to remember that it contains all the compatibilities of a prime implicant.

It may be a spurious extended prime implicant or an impure extended prime implicant or both a spurious and impure extended prime implicant.

Notice that an extended prime implicant can be viewed as a (possibly spurious or impure) mark.

## 5.8. DEFINITION OF REMAINDER

A **PoC** which is neither a (possibly spurious or impure) mark nor an (extended) prime implicant will be called a "remainder". A remainder can be associated to one or more prime implicants, of which it contains a subset of compatibilities.

For example, in the case of **CF**(**4**)

**R = c(2,b;3,c)∗c(2,b;4,d)∗c(3,c;4,d)** **(11)**

is a remainder of the prime implicant **(10)**.

A remainder **R** may be associated to more than one prime implicant. For example, in the case of **CF(3)**, **R=c(2,1;3,1)** is a remainder of the prime implicants

$$\begin{aligned}
\mathbf{P1} &= \mathbf{c(1,1;2,1)*c(1,1;3,1)*c(2,1;3,1)} \\
\mathbf{P2} &= \mathbf{c(1,2;2,1)*c(1,2;3,1)*c(2,1;3,1)} \\
\mathbf{P3} &= \mathbf{c(1,3;2,1)*c(1,3;3,1)*c(2,1;3,1)}
\end{aligned} \qquad \mathbf{(12)}$$

On the definitions of mark and remainder the following properties are based.

### 5.9. Property 4

A not spurious mark **M** specifies a corresponding prime implicant **P** uniquely. Indeed, if all the indexes of triplet appear in **M**, the product **(6)** is completely defined.

We shall write

**P = I(M)**

to state that **P** is the prime implicant specified by **M**.

As already mentioned, a remainder **R** does not specify a corresponding prime implicant uniquely. In the example relative to **CF(3)** above described, three prime implicants correspond to **R = c(2,1;3,1)**, as shown by (**12**), since a single index of triplet is missing in that remainder. In general, if **z** triplets are not involved in **R**, there are $\mathbf{3^z}$ different ways of involving the missing triplets.

Hence the following property follows.

### 5.10. Property 5

A not spurious remainder **R** in which the indexes of **z** triplets are missing corresponds to $\mathbf{3^z}$ different prime implicants.

Finally, the following property can be proved. The proof is not too difficult and it is omitted for the sake of brevity.

### 5.11. Property 6

Let $\mathbf{P_1}$ and $\mathbf{P_2}$ be two **PoC's** such that $\mathbf{P_1 * P_2}$ is equal to a prime implicant **P** of a core function. Either $\mathbf{P_1}$ or $\mathbf{P_2}$ is a mark of **P.**

# 6. THE EXTERNAL CORE FUNCTION

Let $I_j$ be a prime implicant of **CF(n)**. The external core function relative to $I_j$, **ECF(n,$I_j$)**, is defined as the sum of all the minterms of **CF(n)** which imply $I_j$ and no other prime implicant $I_k$ of **CF (n)** with **k≠j**. (Remember that a minterm of a Boolean function **F** is a product of all the variables of **F**, some complemented and some uncomplemented, implying **F**).

Of course,

$$ECF(n,I_j) = I_j * \Pi_{k \neq j} (!I_k) \quad (13)$$

where $!I_k$ denotes the complement of $I_k$, i.e. **(NOT $I_k$).**

The global external core function of order **n,** or **ECF(n)**, will be defined as the sum of **ECF(n, $I_j$)**'s relative to all the prime implicants $I_j$ of **CF(n)**:

$$ECF(n) = \sum_j ECF(n, I_j) \quad (14)$$

The importance of external core function derives from the following analysis.

## 6.1. <u>THEOREM 1</u>

Let **T** be a term (or extended prime implicant) of **CF(n).** It must be the product of all the compatibilities of a prime implicant $I_j$ of **CF(n)** and other compatibilities, that is,

$T = I_j * X$

where **X** is a possibly empty **PoC. T** can also be written as $T = T(I_j)$

All the minterms of **T($I_j$)** contained in **ECF(n)** are minterms of **ECF(n,$I_j$)**.

Indeed, for any **k ≠ j**,

$$T(I_j) * ECF(n,I_k) = I_j * X * I_k * \Pi_{l \neq k} (!I_l) = 0 \quad (15)$$

## 6.2. <u>THEOREM 2</u>

Let **T** be a term of **CF (n)** implying two or more prime implicants of **CF(n)** as, for example,

$T = T(I_j, I_k)$

The number of minterms of

**T($I_j$,$I_k$)** belonging to **ECF(n)** is equal to **0**.

Indeed,

$$T(I_j, I_k) * ECF(n, I_h) = 0 \quad (16)$$

for any **h**.

The preceding theorems 1 and 2 are nearly obvious. On the contrary, the following theorem 3 appears rather complex.

### 6.3. THEOREM 3

Let **T = T ($I_j$) = $I_j$*X** be a term of **CF (n)** which is spurious for a single compatibility **X**.

If **NMT(F)** denotes the number of minterms of Boolean function **F** , the number of minterms of **$I_j$*X** contained in **ECF(n,$I_j$ )** is

$$\mathbf{NMT(Ij * X * ECF(n, Ij))} \;<=\; \frac{\mathbf{1}}{\mathbf{2}} \cdot \mathbf{NMT(ECF(n, Ij))} \quad \mathbf{(17)}$$

The proof of this theorem is presented in Appendix 1.

By proceeding in the same way it is possible to generalize the preceding Theorem 3 as follows.

### 6.4. THEOREM 4

Let

**$I_j * X_1 * X_2 * \ldots X_m$**

are **m s**purious compatibilities.

The number of its minterms contained in **ECF(n, $I_j$)** is

$$\mathbf{NMT(I_j * X_1 * X_2 * \ldots * X_m * ECF(n, I_j))} <= \frac{1}{2^m} \cdot \mathbf{NMT(ECF(n, I_j))} \quad \mathbf{(18)}$$

**Proof**

Also the proof of this theorem is presented in Appendix 1.

The following Theorems 5 and 6 are analogous to preceding Theorems 3 and 4, respectively.

### 6.5. THEOREM 5

Let **T=T ($I_j$)** an impure term of **CF(n)** characterized by a single impure variable **(!X)** :

**T = $I_j$*(!X)**

The number of minterms of **ECF(n,$I_j$)** contained in **T** is

$$\mathbf{NMT}\left(\mathbf{Ij} * (!\,\mathbf{X}) * \mathbf{ECF}(\mathbf{n}, \mathbf{I}_j)\right) \sim \left(\tfrac{1}{2}\right) \cdot \mathbf{NMT}\left(\mathbf{ECF}(\mathbf{n}, \mathbf{I}_j)\right) \quad \mathbf{(19)}$$

See the proof in **Appendix 1.**

### 6.6. THEOREM 6

Let **T=T($I_j$)** an impure term of **CF(n)** characterized by **m** impure variables:

**$T = I_j * (!X_1) * (!X_2) * \ldots (!X_m)$**

The number of minterms of **ECF(n,$I_j$)** contained in **T** is

$$\mathbf{NMT}\left(\mathbf{T} * \mathbf{ECF}(\mathbf{n}, \mathbf{I}_j)\right) \sim \left(\frac{\mathbf{1}}{\mathbf{2}}\right)^{m} \cdot \mathbf{NMT}\left(\mathbf{ECF}(\mathbf{n}, \mathbf{I}_j)\right) \quad \mathbf{(20)}$$

Also Theorem 6 is discussed in Appendix 1.

Notice that **NMT(ECF(n,$I_j$)) = NMT(ECF(n,$I_k$))** for any **j** and **k.** It will be called **NMT1(n)**.

## 7. THE REFERENCE ARCHITECTURE

**Fig. 7** shows the network which will implement core function. By virtue of Theorem 3.1, it does not contain **NOT** gates and it is characterized by a number of subnetworks each of which has the structure shown by **Fig. 8.** As an alternative, the network of **Fig. 7** might be composed by a single network of the type of **Fig. 8**.

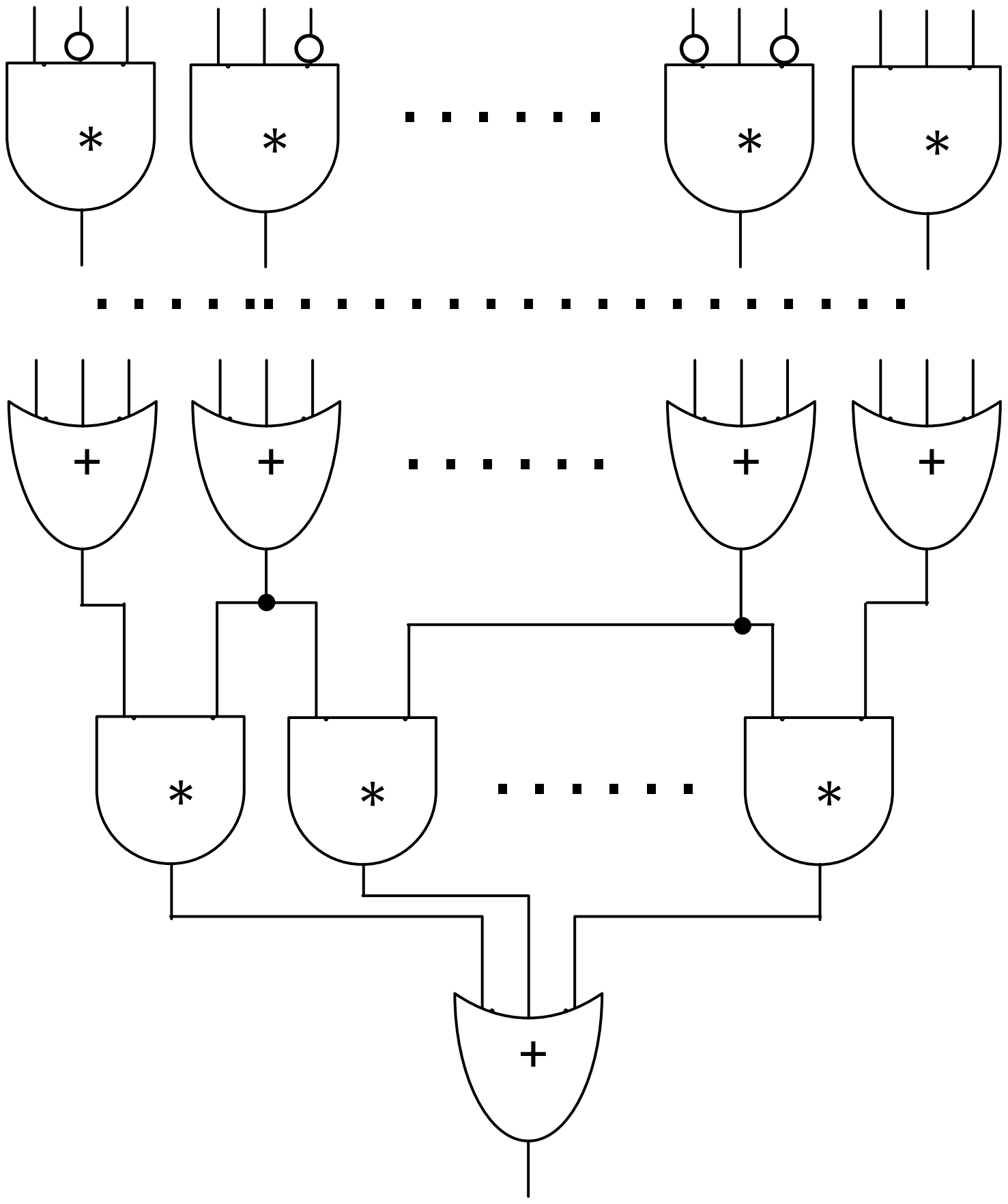

**Fig. 7**

**The Reference Architecture**

The circuit presented in **Fig. 8** will be called a “primary composite addendum (**PCA**)”. Every **$F_i$** will be called a “primary composite addendum factor” (**PCAF**).

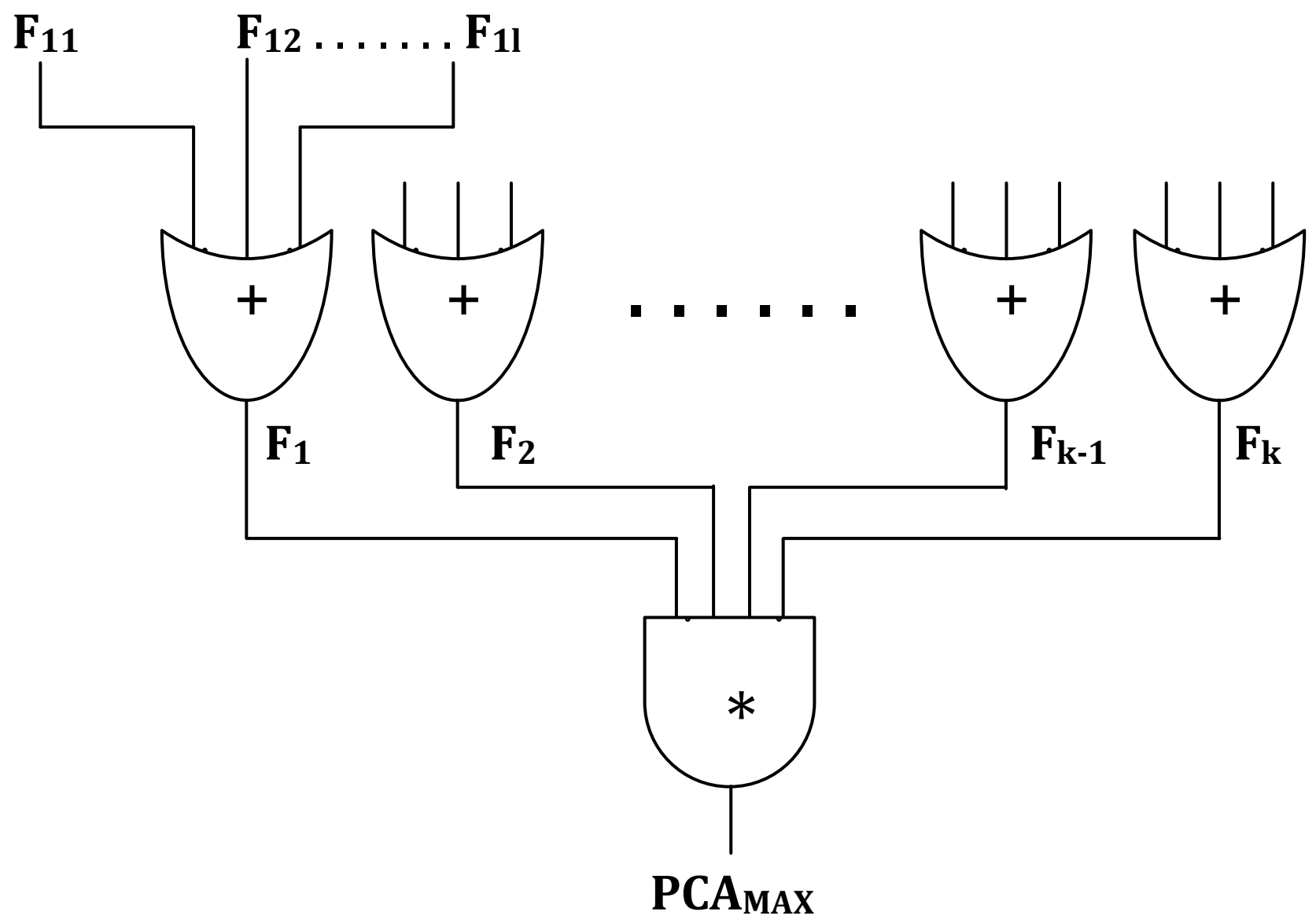

**Fig. 8**

**The primary composite addendum**

If the number of **PCA's** of the minimum cost implementation of **CF(n)** increased with **n** according to an exponential law, also the cost of this implementation would increase according to an exponential law, the cost being represented by the number of **AND** gates at the bottom of **Fig. 7**.

Therefore, the following analysis refers to the case in which the number of **PCA's** of the minimum cost implementation of **CF(n)** increases with **n** according to a polynomial law.

Besides, reference will be made to the following definitions. The merit of a (possibly, impure or spurious) prime implicant **$P_i$** of **CF(n)** will be defined as the number of minterms of **ECF(n)** that **$P_i$** covers and the merit of a **PCA** will be defined as the number of minterms of **ECF(n)** that this **PCA** covers.

We shall discuss the properties of the **PCA** which contains the maximum number of minterms of **ECF(n).** It will be called **$PCA_{MAX}$.**

It is easy to prove that the number of minterms of **ECF(n)** contained in the function implemented by **$PCA_{MAX}$** increases with **n** as **$3^n$**. Besides, also the number of prime implicants of **CF(n)** implemented by **$PCA_{MAX}$** increases with **n** as **$3^n$.**

## 8. SYNTHESIS OF MAXIMUM MERIT PCA

Consider the decomposition of the maximum merit **PCA** into **k** factors

$\mathbf{F_1, F_2, F_3 ..., F_k}$

(**Fig. 8**). Consider also the partial product

$\mathbf{F_{2\text{-}k} = F_2 * F_3 * ... * F_k}$

where the symbols $\mathbf{F_2, F_3, .., F_k}$, above used to denote processing units have the meaning of their corresponding Boolean values, as will be done in the future when such a choice will not generate confusion.

Obviously, the value of the maximum merit **PCA**, that is, the function implemented by it, will be

$\mathbf{val(PCA_{MAX}) = F_1 * F_{2\text{-}k}}$

Let $\mathbf{P_1, P_2, ..., P_v}$ be the prime implicants of function $\mathbf{F_1}$ and $\mathbf{Q_1, Q_2, ..., Q_w}$ the prime implicants of function $\mathbf{F_{2\text{-}k}}$ . Obviously, the value of the maximum merit **PCA** will be the sum of all the $\mathbf{v*w}$ products $\mathbf{P_i*Q_j}$. Some of these products will be equal to **0**; the other ones will be (possibly, impure or spurious) implicants of **CF(n).**

The number of minterms of **ECF(n)** covered by each of these implicants will be defined as its merit.

Notice that any product $\mathbf{P_i*Q_j}$ "must" be an implicant of **CF(n)** (possibly, extended with spurious or impure variables). Otherwise, the considered solution would not be a correct implementation of **CF(n).**

**Fig. 9** shows the symbols which will be used in the following analysis.

An arc connecting node $\mathbf{P_i}$ with node $\mathbf{Q_j}$ denotes that the product $\mathbf{P_i*Q_j}$ is a (possibly impure or spurious) implicant of **CF(t).** For example, this is the case of arcs $\mathbf{P_1 - Q_1}$, $\mathbf{P_1 - Q_2}$, $\mathbf{P_2 - Q_1}$, $\mathbf{P_2 - Q_2}$ in **Fig. 9.** The labels of the arcs $\mathbf{I_0, I_1, I_2, I_3, I_0'}$ (perhaps, the same as $\mathbf{I_0}$), $\mathbf{I_1'}$ are the names of the (possibly extended with spurious or impure variables) prime implicants of **CF(t)** represented by those arcs. A missing arc denotes that the corresponding product is equal to **0**; thus, for example, $\mathbf{P_1*Q_3 = 0}$ or $\mathbf{P_4*Q_3 = 0}$.

Notice that an arc might be labelled with the product of two or more different prime implicants, as in the case of $\mathbf{P_4 - Q_4}$ which has been labelled with the product $\mathbf{I_3*I_4}$. However, as already proved, the merit of the product of two or more different prime implicants is equal to **0.**

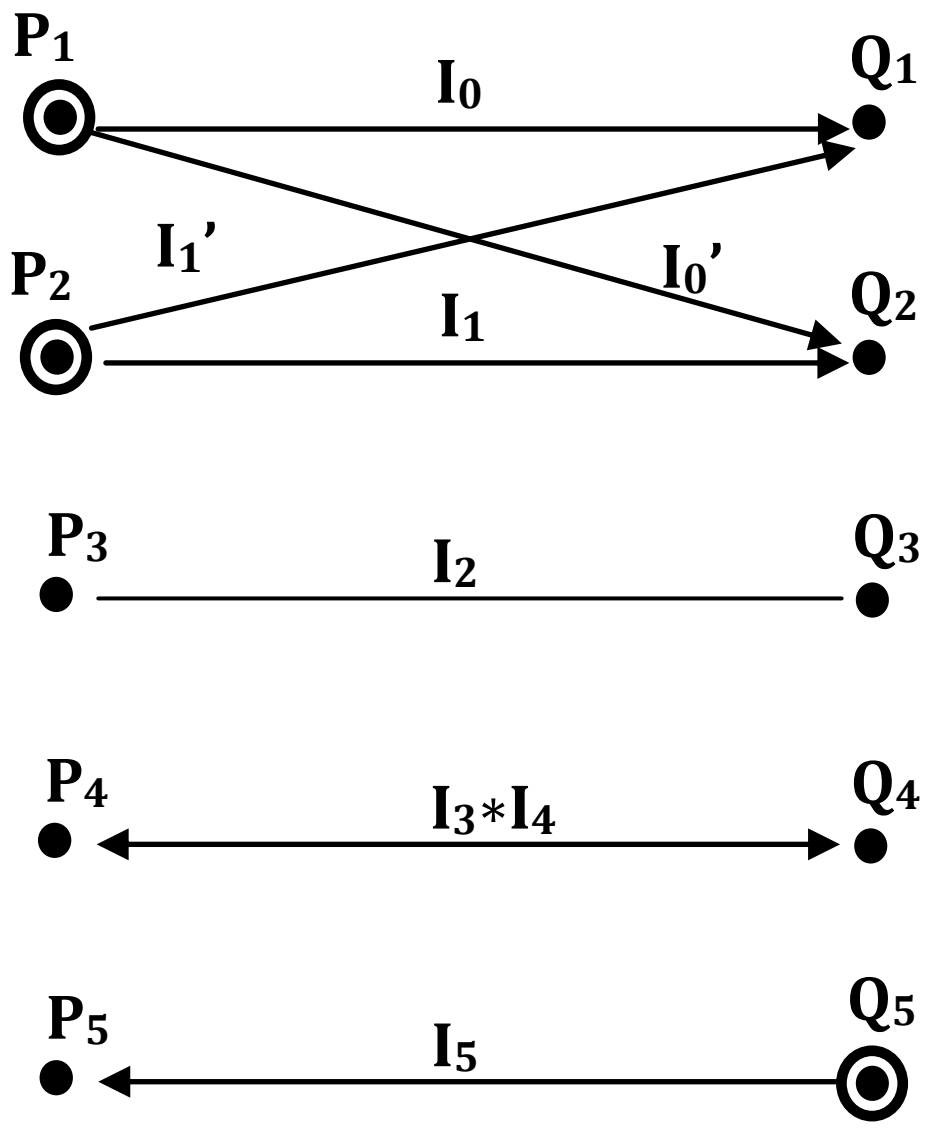

**Fig. 9**

**The Prime Implicants produced by a $PCA_{MAX}$**

Three different cases are worth mentioning.

**<u>Case 1.</u>**

Both $\mathbf{P_3}$ **a**nd $\mathbf{Q_3}$ are marks, and $\mathbf{I(P_3) = I(Q_3)}$. Of course, in this case, $\mathbf{I_2 = I(P_3) = I(Q_3)}$. Notice that, by virtue of Property 6 of previous Section **5**, if $\mathbf{P_3 * Q_3}$ is not equal to **0**, at least one of these two terms is a mark of the generated prime implicant.

**<u>Case 2</u>**

$\mathbf{P_2}$ **i**s a mark and $\mathbf{Q_2}$ is a remainder. Obviously, $\mathbf{I_1 = I(P_2)}$. The considered arc is oriented from $\mathbf{P_2}$ to $\mathbf{Q_2}$ in order to remember that $\mathbf{P_2}$ is the "origin" of the arc, that is, the mark of the corresponding prime implicant.

This is also the case of the arcs $\mathbf{P_1 - Q_1}$, $\mathbf{P_1 - Q_2}$, $\mathbf{P_2 - Q_1}$.

Notice that in **Case 1** both $\mathbf{P_3}$ and $\mathbf{Q_3}$ might be considered as origins of the prime implicant $\mathbf{I_2 = P_3 * Q_3}$.

**<u>Case 3</u>**

$\mathbf{P_5}$ is a mark of a prime implicant $\mathbf{I(P_5)}$ while $\mathbf{Q_5}$ is a mark of a different prime implicant $\mathbf{I(Q_5) \neq I(P_5)}$. Since the produced prime implicant $\mathbf{I_5}$ coincides with $\mathbf{I(Q_5)}$, the arc has been oriented from $\mathbf{Q_5}$ which is considered as the origin of the arc.

- - - - - - - - -

Since the number of implicants implemented by $\mathbf{PCA_{MAX}}$ increases with **n** as $\mathbf{3^n}$, also the number of origins born in the decomposition of **Fig. 5** increases with **n** as $\mathbf{3^n}$.

Assume that the number of origins labeled as $\mathbf{Q_j}$ is larger than the number of origins labeled as $\mathbf{P_i}$. In this case, of course, the number of $\mathbf{Q_j}$ origins increases with **n** as $\mathbf{3^n}$. The case in which the number of $\mathbf{Q_j}$ origins is less than the number of $\mathbf{P_i}$ origins can be treated in a similar way.

# 9 THE REDUCTIONS IN $\mathbf{PCA_{MAX}}$

As shown in previous sections $\mathbf{PCA_{MAX}}$ is a subsystem of **CF(n)** covering a number of minterms of **ECF(n)** of the order of $\mathbf{3^n \cdot NMT1(n)}$. In this case

$\mathbf{F_{2\text{-}k} = Q_1+Q_2+\ldots+Q_z}$

where **z** increase with **n** as $\mathbf{3^n}$ and each $\mathbf{Q_i}$ is a mark.

Now consider the decomposition

$\mathbf{F_{2\text{-}k} = F_2 * F_{3\text{-}k}}$

where $\mathbf{F_2}$ and $\mathbf{F_{3\text{-}k}}$ can be written as sums of their prime implicants

$\mathbf{F_2 = R_1+R_2+\ldots}$

$\mathbf{F_{3\text{-}k} = S_1+S_2+\ldots}$

This decomposition makes it possible to reduce the number of involved marks by one unit as follows:

$\mathbf{F_{2\text{-}k} = Q_1+Q_2+Q_3}$

$\mathbf{F_2 = R_1+Q_2+Q_3}$

$\mathbf{F_{3\text{-}K} = S_1+Q_2+Q_3}$

where mark $\mathbf{Q_1}$ has been decomposed into the product $\mathbf{R_1 * S_2}$ of the two remainders $\mathbf{R_1}$ and $\mathbf{S_1}$.

For example, consider the following functions relative to **CF(3)**:

$\mathbf{P_i = c(1,1;2,1) * c(1,1;3,1) * c(2,1;3,1)}$

$\mathbf{F_{2\text{-}k} = Q_1+Q_2+Q_3}$

where

$\mathbf{Q_1 = c(1,1;4,1) * c(2,1;4,1) * c(3,1;4,1)}$

$\mathbf{Q_2 = c(1,1;4,2) * c(2,1;4,2) * c(3,1;4,2)}$

$\mathbf{Q_3 = c(1,1{:}4,3) * c(2,1;4,3) * c\ (3,1;4,3)}$

In this case

$\mathbf{F_{2\text{-}k} = F_2 * F_{3\text{-}k}}$

where

$\mathbf{F_2 = c(1,1;4,1) * c(\ 2,1;4,1) + Q_2 + Q_3}$

$\mathbf{F_{3\text{-}k} = c(3,1;4,1) + Q_2 + Q_3}$

We can define the "merit" $\boldsymbol{\mu}$ of a transformation of the type of the preceding one as the number of minterms of **ECF(n)** covered by the marks $\mathbf{Q_j}$ of $\mathbf{F_{2\text{-}k}}$ which did not appear in $\mathbf{F_2}$ and $F_{3\text{-}k}$. Of course, in the case of previous example and in all the cases in which a single $Q_i$ is decomposed into two remainders,

$$\mu = NMT1(n)$$

In this example all involved compatibilities are pure and not spurious. In this section we restrict our attention to this hypothesis; the case of impure or spurious compatibilities will be discussed in next section.

It is easy to prove that both $F_2$ and $F_{3\text{-}k}$ must contain at least **(z-1)** marks, that is, in the best case, the decomposition of $F_{2\text{-}k}$ produces the reduction by one unit of the number of marks contained in $F_{2\text{-}k}$. In other terms, the product of $F_2$ by $F_{3\text{-}k}$ produces a new mark in the best case.

Indeed, assume, for example, that

$$R_1 * S_1 = Q_1$$

$$R_2 * S_2 = Q_2$$

where $R_1$, $R_2$, $S_1$ and $S_2$ are all remainders. It is easy to verify that if $R_1$, $R_2$, $S_1$ and $S_2$ are all remainders, $R_1 * S_2 * P_i$ and $R_2 * S_1 * P_i$ are not implicants of **CF(n).**

Indeed, for example, if **CF(5)** is the considered core function and

$$P_i=c(1,1;2,1)*c(1,1;3,1)*c(2,1;3,1)$$

$$Q_1=c(1,1;4,1)*c(1,1;5,1)*c(2,1;4,1)*c(2,1;5,1)*c(3,1;4,1)*c(3,1;5,1)*c(4,1;5,1)$$

$$Q_2=c(1,1;4,2)*c(1,1;5,1)*c(2,1;4,2)*c(2,1;5,1)*c(3,1;4,2)*c(3,1;5,1)*c(4,2;5,1)$$

$$R_1=c(1,1;4,1)*c(1,1;5,1)*c(2,1;4,1)*c(2,1;5,1)*c(4,1;5,1)$$

*(which is a remainder since* **<3,1>** *is missing)*

$$S_1=c(2,1;4,1)*c(2,1;5,1)*c(3,1;4,1)*c(3,1;5,1)*c(4,1;5,1)$$

*(which is a remainder since* **<1,1>** *is missing)* **(21)**

$$R_2=c(1,1;4,2)*c(1,1;5,1)*c(2,1;4,2)*c(2,1;5,1)*c(4,2;5,1)$$

*(which is a remainder since* **<3,1>** *is missing)*

$$S_2=c(2,1;4,2)*c(2,1;5,1)*c(3,1;4,2)*c(3,1;5,1)*c(4,2;5,1)$$

*(which is a remainder since* **<1,1>** *is missing)*

The product $P_i * R_1 * S_2$ does not imply $P_i * Q_1$ since **c(3,1;4,1)** is missing and it does not imply $P_1 * Q_2$ since **c(1,1;4,2)** is missing, while the product $P_i * R_2 * S_1$ does not imply $P_i * Q_1$ since **c(1,1;4,1)** is missing and it does not imply $P_i * Q_2$ since **c(3,1;4,2)** is missing.

In more general terms, the proof of this property can be stated as follows.

Since **PoC** $R_1$ is a remainder, at least one of the indexes characterizing $Q_1$ does not appear in $R_1$ (as index **<3,1>** in the above example). Let it be **<p,q>**.

If $Q_1$ is different from $Q_2$, $Q_1$ must contain at least one index which does not appear in $Q_2$ (index **<4,1>** in the above example). Let it be **<r,s>**.

It follows that the compatibility **<p.q; r,s>**, which is necessary in order that the product $R_1 \star S_2$ implies $Q_1$ does not appear in $R_1 \star S_2$.

For a similar reason, $\mathbf{R_1} \star \mathbf{S_2}$ does not imply $\mathbf{Q_2}$ and all the other implicants of the implemented function.

## 10. DECOMPOSITIONS INVOLVING IMPURE OR SPURIOUS PRODUCTS OF COMPATIBILITIES

If the condition of involving only pure and not spurious **P** o **C** is removed, a single decomposition can make it possible to reduce the number of marks in the sum **($Q_1$+$Q_2$+....)** by two or more than two units.

The most interesting example is the following one relative to **CF (4).**

With

**$P_i$ = c(1,1;2,1)*c(1,1;3,1)*c(2,1;3,1),**

**$Q_1$+$Q_2$+$Q_3$+$Q_4$+……=**

**c(1,1;4,1)*c(2,1;4,1)*c(3,1;4,1)+**

**c(1,1;4,2)*c(2,1;4,2)*c(3,1;4,2)+**

**$Q_3$+$Q_4$+……=**

**(c(1,1;4,1)*c(2,1;4,1)+c(1,1;4,2)*c(2,1;4,2)+$Q_3$+$Q_4$+…)*** **(22)**

**(c(3,1;4,1)*c(2,1;4,1)+c(3,1;4,1)*c(3,1;4,2)+$Q_3$+$Q_4$+…)**

This decomposition produces the reduction of the pure and not spurious mark

**c(1,1;4,1)*c(2,1;4,1)*c(3,1;4,1)**

and of the spurious mark

**c(1,1;4,2)*c(2,1;4,2)*c(3,1;4,2)*c(1,1;4,1)*c(3,1;4,1)**

The merit associated to the first mark is

**μ(c(1,1;4,1)*c(2,1;4,1)*c(3,1;4,1)) = NMT1 (4),**

while, according to Theorem 4 of section 6, the merit associated to the second mark is

**μ(c(1,1;4,2)*c(2,1;4,2)*c(3,1;4,2)*c(1,1;4,1)*c(3,1;4,1)) = 1/4*NMT1** (4)

since this mark contains two spurious compatibilities **(c(1,1;4,1)** and **c(3,1;4,1))**.

It follows that the total merit of this reduction is **1.25*NMT1 (4)**.

This reduction is an exception in the framework of multiple reductions since its merit is (slightly) larger than **NMT1 (n)**.

Indeed, the merit of a multiple reduction is a decreasing function of the number **NRM** of reduced marks and of the number **NDV** of variables which are different from a mark and another one.

The best decompositions for various values of **NRM** and **NDV** and the corresponding values of total merit are presented in **Appendix 2.** In the same appendix also the reasons for which the total merit decreases when the number of reduced marks increases are briefly discussed.

## 11. DECOMPOSITIONS AT HIGHER LEVELS

The product $F_1*F_2*...*F_k$ may produce other marks in addition to those generated inherently by the product $F_1*F_2*...*F_k$. Indeed, one or more marks of **CF(n)** can be implicants of some $F_j$.

For example, consider function $F_1$ implemented by the first of **PCAF's** represented in **Fig. 10.**

$F_1$ is the output of an **OR** gate. Indeed, if it were the output of an **AND** gate, this might be merged together with the **AND** gate producing the output of the considered $PCA_{MAX}$ with the reduction of the cost by one unit.

Let $F_{11}, F_{12}, ..., F_{1l}$ be the inputs of this **OR** gate (**Fig. 10**). In its turn, node $F_{11}$ contains a mark or a sum of marks as the product of functions $F_{111}, F_{112}, F_{113}, ...$

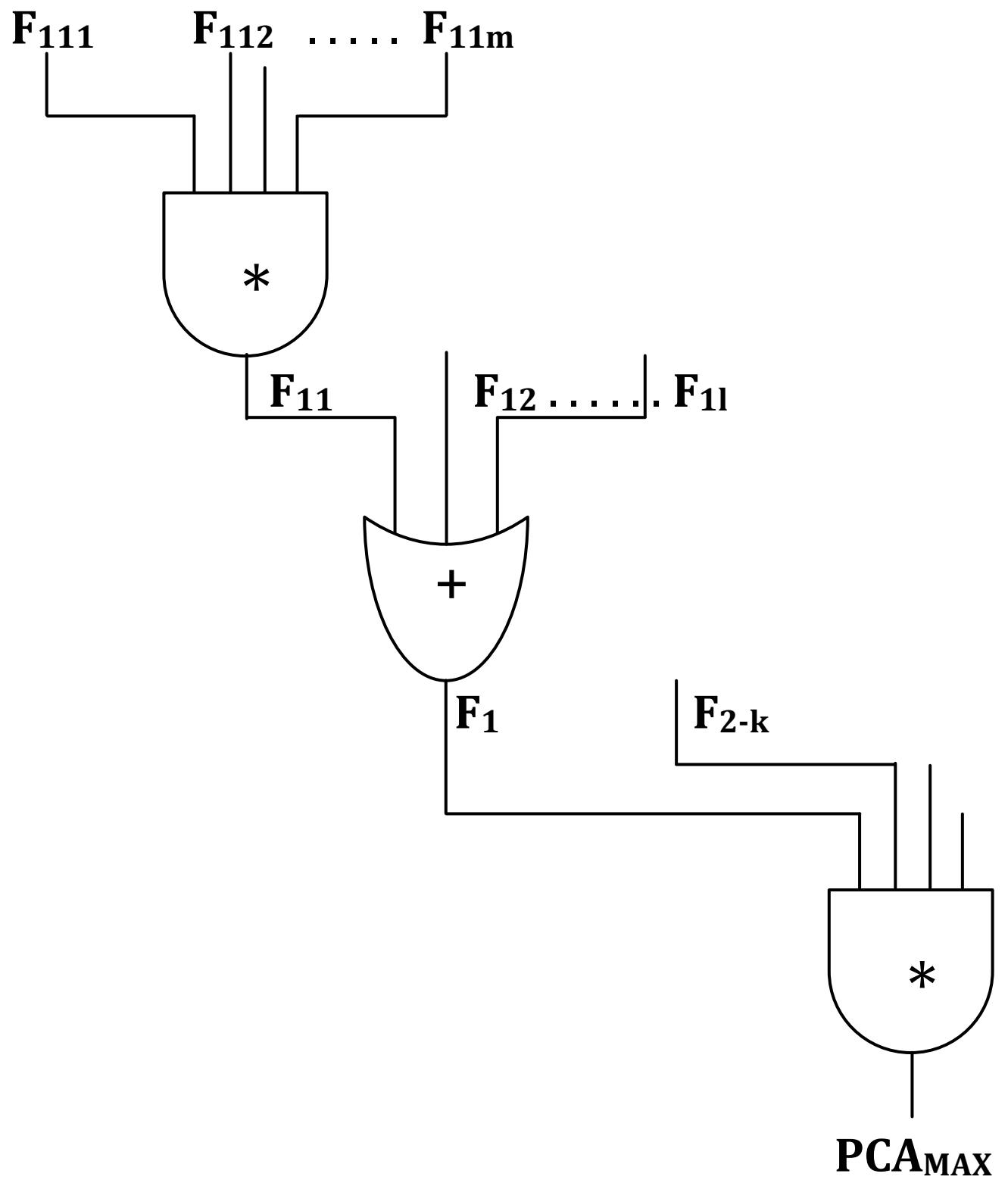

**Fig. 10**

**The decomposition of primary composite addenda**

Indeed, it is easy to prove that a mark available on $F_1$ cannot be the sum of two or more than two remainders available on two or more than two nodes $F_{1j}$ and $F_{1k}$. For example, the mark

**c(1,1;2,1) ⋆ c(1,1;3,1) ⋆ c(2,1;4,1)**

might be the sum of the two remainders

**c(1,1;2,1) ⋆ c(1,1;3,1) ⋆ c(2,1;3,2)** (24)

and

**c(2,1;4,1) ⋆ !c(2,1;3,2)**

of **CF(4)**, available in $F_{1j}$ and $F_{1k}$, but these two remainders would be prime implicants of $F_1$ and, therefore, $PCA_{MAX}$ would not imply **CF(4)** for the reasons discussed in **Section 8**.

If an $F_{1j}$ is a mark, it can be implemented as the **AND** of all its compatibilities with a merit **μ=NMT(n)**.

If an $F_{1j}$ is a sum of marks, it can be decomposed into products of sums as described in **Sections 8** and **9**. Each elementary decomposition has a merit of the order of **NMT1(n)** but it has the cost of at least one gate.

The considerations above developed on the **AND** gate producing $PCA_{MAX}$ (at the bottom of **Fig. 10**) apply also to the **AND** gate producing $F_{11}$ and to all the **AND** gates introduced at the higher levels. Each of these **AND** gates can generate marks at the cost of a number of gates equal to the total merit of these marks divided by **NMT1(n)**.

In order to simplify this statement we can state that, in a first approximation, the **AND** gate of **Fig. 10** producing $PCA_{MAX}$ can generate **h** marks at the cost of **h** gates, that is, the **OR** gates producing $F_{111}$, $F_{112}$,...This result apply to all the **AND** gates used in the implementation of $PCA_{Max}$.

The considerations above developed on the **AND** gates producing $PCA_{MAX}$ (at the bottom of **Fig. 10**) apply also to the **AND** gate producing $F_{11}$. This gate can generate **h-1** origins at the cost of **h** gates , that is, the **OR** gates producing $F_{111}$, $F_{112}$, $F_{113}$...

## 12. CONCLUSIONS

Consider an **AND** gate of the network $N_M$ implementing Core Function with the minimum number of gates . For the sake of simplicity, assume that such a gate has only two inputs **A** and **B** and one output **C**; indeed, an **AND** gate characterized by **i** inputs can be decomposed into **(i-1) AND** gates each of which has two inputs only.

Let us define the merit of a function **f(m)** available in a node **m** of $N_M$ as follows.

Write **f(m)** as the sum of its prime implicants. The merit of **f(m)** is defined as the number of minterms of **ECF(n)** contained in all the prime implicants of **f(m)** which are marks of Core Function.

As quickly stated in Section **9**, the merit of the considered **AND** gate will be defined as the number of minterms of **ECF(n)** covered by the marks $Q_j$ of output **C** which did not appear in input **B** or in input **C.** It is easy to prove that the merit of the considered gate is equal to the difference

**merit (f(C)) – merit (f(A)+f(B))**

For example, in the case of Eq.**21** of Section **9** characterized by the following values:

**$f(A) = F_2$**

**$f(B) = F_{3-k}$**

**$f(C) = F_{2-k}$**

we can write:

**merit (f(A)) = 2*NMT1(n)**

**merit (f(B)) = 2*NMT1(n)**

**merit (f(A) + f(B)) = 2*NMT1(n)**

**merit (f(C)) = 3*NMT1 (n)**

Therefore, the merit of the **AND** gate performing the product **(A)*(B)** is equal to **NMT1(n).**

Similar considerations can be applied to **AND** gates involving impure or spurious products of shown in Section **10** and related appendixes. Even in the best cases the merit of a gate is always of the order of **NMT1(n).**

Now consider an **OR** gate characterized by two inputs **A** and **B** and an output **C**.

In this case the equation

**merit of the gate = merit (f(C)) – merit (f(A)+f(B))**

is not true as shown by the example **(24)** of Section **11**, according which the merit of the gate would be equal to **1.** However, the two remainders producing the considered mark would make the implementation of the Core Function no more valid, as shown by the same example of Section **11.** It follows that the merit of an **OR** is always equal to **0.**

Since the total merit of the best network implementing Core Function is equal to **$3^n$*NMT1 (n),** the merit of an **AND** gate is of the order of **NMT1(n)** and the merit of an **OR** gate is equal to **0,** the number of gates contained in the considered network must be of the order of **$3^n$**, at least.

Since the synthesis of core function is an **NP-**complete problem, this result is equivalent to proving that **P** and **NP** do not coincide.

# APPENDIX 1

PROOF OF THEOREM 6.3

For the sake of simplicity, without any loss of generality, assume:

**n = 3**

**X = c(1,1;2,2)**

**$I_1$ = c(1,1;2,1)∗c(1,1;3,1)∗c(2,1;3,1).**

In this case

**ECF(3,$I_1$) = $I_1$∗(!$I_2$)∗(!$I_3$)∗...∗(!$I_{27}$)**

where

**!$I_2$ = !c(1,1;2,1)+!c(1,1;3,2)+!c(2,1;3,2)**

**!$I_3$ = !c(1,1;2,1)+!c(1,1;3,3)+!c(2,1;3,3)**

$!I_4 = !c(1,1;2,2)+!c(1,1;3,1)+!c(2,2:3,1)$

$!I_5 = !c(1,1;2,2)+!c(1,1;3,2)+!c(2,2;3,2)$

$!I_6 = !c(1,1;2,2)+!c(1,1;3,3)+!c(2,2;3,3)$

$!I_7 = !c(1,1;2,3)+!c(1,1;3,1)+!c(2,3;3,1)$

$!I_8 = !c(1,1;2,3)+!c(1,1;3,2)+!c(2,3;3,2)$

$!I_9 = !c(1,1;2,3)+!c(1,1;3,3)+!c(2,3;3,3)$

$!I_{10} = !c(1,2;2,1)+!c(1,2;3,1)+!c(2,1;3,1)$

.

.

.

$!I_{27} = !c(1,3;2,3)+!c(1,3;3,3)+!c(2,3;3,3)$

Consider the two following functions:

$H =$

$+!c(2,2;3,1)*!c( 1,1:3,2)*!c(1,1;3,3)+$

$+ !c(2,2;3,1)*!c(1,1;3,2)*!c(2,2;3,3)+$

$+!c(2,2;3,1)*!c(2,2;3,2)*!c(1,1;3,3)+$

$+!c(2,2;3,1)*!c(2,2;3,2)*!c(2,2;3,3)$

and

$F = I_1*(!I_2)*(!I_3)*(!I_7)*(!I_8)*...(!I_{27})$

from which the following equation derives:

$ECF(3,I_1) = F*!c(1,1;2,2)+F*H$

where **F** and **H** do not contain variable **c(1,1;2,2).**

Function $ECF(3,I_1)$ is the sum of the three following functions:

$F_1 = !c(1,1;2,2)*F*(!H)$

$F_2 = !c(1,1;2,2)*F*H$

$F_3 = c(1,1;2,2)*F*H$

These functions are disjoint in the sense that $F_i$ contains none of the minterms contained in $F_j$ with $j <> i$ . Therefore,

$NMT(ECF(3,I_1)) = NMT(F_1) + NMT(F_2) + NMT(F_3)$

and

$NMT(F_2) = NMT(F_3)$ ,

from which

$NMT(c(1,1;2,2)*I_1* ECF(3,I_1)) = NMT(F_3) < ½*NMT(ECF(3,I_1))$

PROOF OF THEOREM 6.4

For the sake of brevity, the proof is restricted to the case **m=2**.

In this case we can write:

$$ECF(n, I_j) = I_j*A*(!X_1)+I_j*B*(!X_2)+I_j*C*(!X_1)*(!X_2)+I_j*D$$

where functions **A, B, C, D** do not contain variables $X_1$ or $X_2$.

Notice that

$$X_1*X_2*ECF(n, I_j) = X_1*X_2*I_j*D$$

and $(X_1*X_2*D)$ contains **¼** of the minterms of **D.**

Following the same line of reasoning followed in the proof of **Theorem 6.3,** it is easy to prove that

$$NMT(I_j*X_1*X_2*ECF(n, I_j)) <= 1/4 \cdot NMT(ECF(n, I_j))$$

PROOF OF THEOREM 6.5

For the sake of simplicity, without any loss of generality, assume again:

$$n = 3$$

$$X = c(1,1;2,2)$$

$$I_1 = c(1,1;2,1)*c(1,1;3,1)*c(2,1;3,1).$$

Consider again the following functions defined in the proof of **Theorem 6.3** :

$$F_1 = !c(1,1;2,2)*F*(!H)$$

$$F_2 = !c(1,1;2,2)*F*H$$

$$F_3 = c(1,1;2,2)*F*H$$

It is apparent that

$$NMT(!c(1,1;2,2)*I_1* ECF(3,I_1)) = NMT(F_1) + NMT(F_2)$$

Consider a prime implicant **U** of function F containing neither **!c(1,1;3,2)** nor **!c(2,2;3,2).** This prime implicant **U**, multiplied by **V = c(1,1;3,2)*c(2,2;3,2)**, which is a prime implicant of function **(!H)** , produces a **PoC U*V** . All the minterms implying **U*V** are minterms of function $F_1$. For any minterm of function $F_1$ there exists a minterm of function $F_2$ because **U*!c(1,1;3,2)*!c(2,2;3,2)** implies $F_2$ .

But a prime implicant **U** of **F** containing **!c(1,1;3,2)** or **!c(2,2;3,2)** multiplied by V is equal to **0** while the same prime implicant **U** of **F** multiplied by **V** produces an implicant of $F_2$ containing many minterms.

Therefore, $NMT(F_1)$ is smaller than $NMT(F_2)$, and therefore, since $NMT(F_2)$ is equal to $NMT(F_3)$,

$$1/2 < NMT(F_2) = NMT(!c(1,1;2,2)*I_1* ECF(3,I_1)) < 2/3$$

Since it is easy to verify that the number of products of prime implicants of **F** by prime implicants of **(!H)** equal to **0** is very large we can write:

$$NMT(!c(1,1;2,2)*I_1* ECF(3,I_1)) \sim 1/2$$

PROOF OF THEOREM 6. 6

The proof is left to the reader since it can be obtained by applying the methods used in the proofs of **Theorem 6.4** and **Theorem 6.5.**

## APPENDIX 2

THE BEST IMPURE OR SPURIOUS DECOMPOSITIONS

**NRM = 2 NDV = 1**

**$P_1$ = c(1,1;2,1)*c(1,1;3,1)*c(2,1;3,1);**

**$Q_{11}$ = c(1,1;4,1)*c(2,1;4,1);**

**$Q_{21}$ = (3,1;4,1)*c(3,2;4,1);**

**$P_2$ = c(1,1;2,1)*c(1,1;3,2)*c(2,1;3,2);**

**$Q_{12}$ = c(1,1;4,2)*c(2,1;4,1);**

**$Q_{22}$ = c(3,1;4,1)*c(3,2;4,1)**

**μ = 1**

**-----------------**

**NRM = 2 NDV = 1**

**$P_1$ = c(1,1;2,1)*c(1,1;3,1)*c(2,1;3,1)**

**$Q_{11}$ = !c(1,2;2,1)*c(1,1;4,1);**

**$Q_{21}$ = c(2,1;4,1)*c(3,1;4,1)**

**$P_2$ = c(1,2;2,1)*c(1,2;3,1)*c(2,1;3,1)**

**$Q_{12}$ = !c(1,1;2,1)*c(1,2;4,1)**

**$Q_{22}$ = $Q_{21}$**

**μ = 1**

-------------

**NRM = 2 NDV = 2**

**$P_1$ = c( 1,1;2,1)*c(1,1;3,1)*c(2,1;3,1);**

**$Q_{11}$ = c(1,1;4,1)*c(2,1;4,1)*c(2,2;4,1)**

**$Q_{21}$ =c(3,1;4,1)*c(3,2;4,1)**

**$P_2$ = c(1,1;2,2)*c(1,1;3,2)*c(2,2;3,2)**

**$Q_{12} = Q_{11}$ ;    $Q_{22} = Q_{21}$**

**μ = 0,5**

**--------------**

**NRM = 3        NDV = 1**

**$P_1$ = c(1,1;2,1)*c(1,1;3,1)*c(2,1;3,1)**

**$Q_{11}$ = c(1,1;4,1)*c(2,1;4,1)**

**$Q_{21}$ = c(3,1;4,1)*c(3,1;4,2)*c(3,1;4,3)**

**$Q_{12}$ = c(1,1;4,2)*c(2,1;4,2)**

**$Q_{22} = Q_{21}$**

**$Q_{13}$ = c(1,1;4,3)*c(2,1;4,3)**

**$Q_{23} = Q_{11}$**

**μ = 0,75 * NMT1(4)**

THE REASONS FOR WHICH THE MERIT DECREASES WHEN **NRM** INCREASES

Assume that a given decomposition makes it possible to reduce **N** marks with merit equal to **M** and that we want to reduce an **(N+1)-th** mark with that decomposition.

Each product $P_i*Q_{i1}*Q_{i2}$ of the **N** lines of a previous decomposition must be corrected by adding a suitable compatibility or product of compatibilities in order that, for example, $P_i*Q_{i1}*Q_{j2}$ (where $Q_{j2}$ belongs to the **(N+1)-th** line) is an implicant of Core Function.

This correction implies the multiplication of the merit **M** by $1/(2^N)$ and therefore a reduction equal to $M-M/(2^N)$.

The new line must be compatible with the **N** preceding lines and therefore its merit is of the order of $NMT1(n)/(2^N)$, whose absolute value is much less than **M.**